\documentclass{WileyMSP-template}

\usepackage{siunitx}
\usepackage{float}
\usepackage{subfigure}

\begin{document}

\pagestyle{fancy}
\rhead{\includegraphics[width=2.5cm]{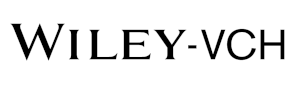}}

\title{Nonlocal Differential Resistance in AlO$_x$/KTaO$_3$ Heterostructures}

\maketitle


\author{Patrick. W. Krantz},
\author{Venkat Chandrasekhar*}

\dedication{}

\begin{affiliations}
Northwestern University\\
2145 Sheridan Rd.\\
Evanston, Illinois, US, 60208\\
Email Address: v-chandrasekhar@northwestern.edu

\end{affiliations}


\keywords{Complex Oxides, Cryogenic, Differential Resistance}

\begin{abstract}

Local and nonlocal differential resistance measurements on Hall bars defined in AlO$_x$/KTaO$_3$ heterostructures show anomalous behavior that depends on the crystal orientation and the applied back gate voltage.  The local differential resistance is asymmetric in the dc bias current, with an antisymmetric component that grows with decreasing gate voltage.  More surprisingly, a large nonlocal differential resistance is observed that extends between measurement probes that are separated by 100s of microns.  The potential source of this anomalous behavior is discussed.

\end{abstract}


\paragraph{}
Significant attention has been paid to devices built on complex oxides for nearly two decades, and for good reason. Since the advent of field, sparked by the discovery of conductivity between resistive layers of LaAlO$_3$ deposited on SrTiO$_3$ (STO) substrates \cite{ohtomo_high-mobility_2004}, a number of properties have been shown to live in the two dimensional electron gas (2DEG) at the interfaces of these complex oxide heterostructures. These properties include superconductivity \cite{reyren_superconducting_2007, han_two-dimensional_2014, monteiro_two-dimensional_2017}, strong spin-orbit interactions \cite{caviglia_tunable_2010}, and magnetism \cite{brinkman_magnetic_2007, dikin_coexistence_2011}, and have often been shown to coexist on the same sample \cite{dikin_coexistence_2011, li_coexistence_2011}. Importantly, these properties were shown to be tunable with external parameters, such as an applied back gate voltage, $V_g$ \cite{caviglia_tunable_2010, thiel_tunable_2006, bell_dominant_2009}, generating a platform for further device development. Recently, focus has shifted to a new complex oxide, KTaO$_3$ (KTO), which is similar to STO in some ways, but also has some significant differences. For example, like STO based heterostructures, KTO structures go superconducting but at significantly higher temperatures \cite{liu_two_dimensional_2021, chen_two-dimensional_2021, qiao_gate_2021}. Another significant difference is that KTO has a much larger spin orbit coupling and a complex, crystal direction dependent spin texture\cite{bruno_band_2019}. One consequence of the large spin orbit interactions is the possibility of efficient spin-charge conversion \cite{vicentearche_spincharge_2021}, making KTO a promising platform for spintronic applications \cite{gupta_ktao_2022}.

\paragraph{}
We report here anomalous low temperature local and nonlocal differential resistance measurements in 2DEGs that form at the interface between AlO$_x$ and KTO in the normal state.  The local differential resistance is asymmetric in the applied dc bias current, with the asymmetry increasing as the back gate voltage $V_g$ is decreased.  A large nonlocal differential resistance is observed that also increases with decreasing $V_g$, and extends over lengths on the order of 100s of microns.  The nonlocal differential resistance depends on the specific probe configuration used to measure it, and is greatly reduced or almost completely disappears in certain configurations.  This feature of the nonlocal signal rules out thermal effects, and suggests instead a chiral nature of charge transport as the source of the anomalous behavior.  


\paragraph{}
The KTaO$_3$ substrates used in this study were 5 mm x 5 mm x 0.5 mm single side polished crystals purchased from MSE Supplies LLC. Crystals with 3 different surface orientations ((001), (110) and (111)) were processed identically, first through a standard cleaning regimen of 3 min ultrasonication in acetone, 3 min ultrasonication in DI water and 3 min ultrasonication in isopropanol, then through photolithography patterning using LOR-5A and S-1813 photoresists. The photolithography process defined Hall bars that were 50 $\mu$m wide with 600 $\mu$m between sets of voltage probes which had a 2 $\mu$m constriction defining the contact point. The devices were realized using a series of 99.9995\% aluminum depositions with an electron-gun evaporator system. The deposited Al getters oxygen from the KTO surface, creating oxygen vacancies that result in a conducting 2DEG, a procedure which was first developed for generating 2DEGs on STO substrates \cite{delahaye_metallicity_2012}. The processing steps for these depositions were as follows:  after an in-situ oxygen plasma clean to remove any residual photoresist, a layer of 1.5 nm of aluminum was deposited onto the KTO substrates and allowed to sit for 10 min before several more 1.5 nm layers were deposited to protect the device. The second and subsequent layers were completely oxidized by introducing 100 mTorr of oxygen into the chamber for 10 min before the application of the next layer. The result was a resistive, amorphous AlO$_x$ over-layer and a 2DEG formed by the oxygen vacancies on the surface \cite{mallik_superfluid_2022}. Samples were then wirebonded to a header and affixed at the bottom of a Oxford MX100 cryostat equipped with a two axis superconducting magnet. 

\paragraph{}
Differential resistance measurements were conducted by summing a low frequency sine wave from a lock-in amplifier ($<$20 Hz) with a dc voltage from an Agilent 33500B waveform generator.  The output of the summer was then used to drive a custom-built current source.  Typical ac currents were 20 nA, and for all measurements the dc and ac currents were applied along the same path. The voltage output from the devices was amplified by a first stage INA110 based pre-amplifier before being read by the lock-ins for the ac signal, and a HP34401A digital multimeter for the dc signal. The back gate voltage $V_{g}$ was supplied by a Keithley KT2400 source equipped with a low pass filter and measured independently by another HP34401A digital multimeter. Three different measurement configurations based on Hall bar geometries were measured at $\sim$ 5K in these experiments, which are discussed in more detail below. It is important to note that the resistance of these devices is quite high when compared to similar geometries constructed from metals or cleaner 2DEGs \cite{jin_nonlocal_2017, eom_asymmetric_1996}. Thus, the scale of the sample resistance limits the magnitude of the applied dc current I$_{dc}$ to 1 $\mu$A in this study.

\begin{figure}[h]
    \centering
    \includegraphics[width=0.6\columnwidth]{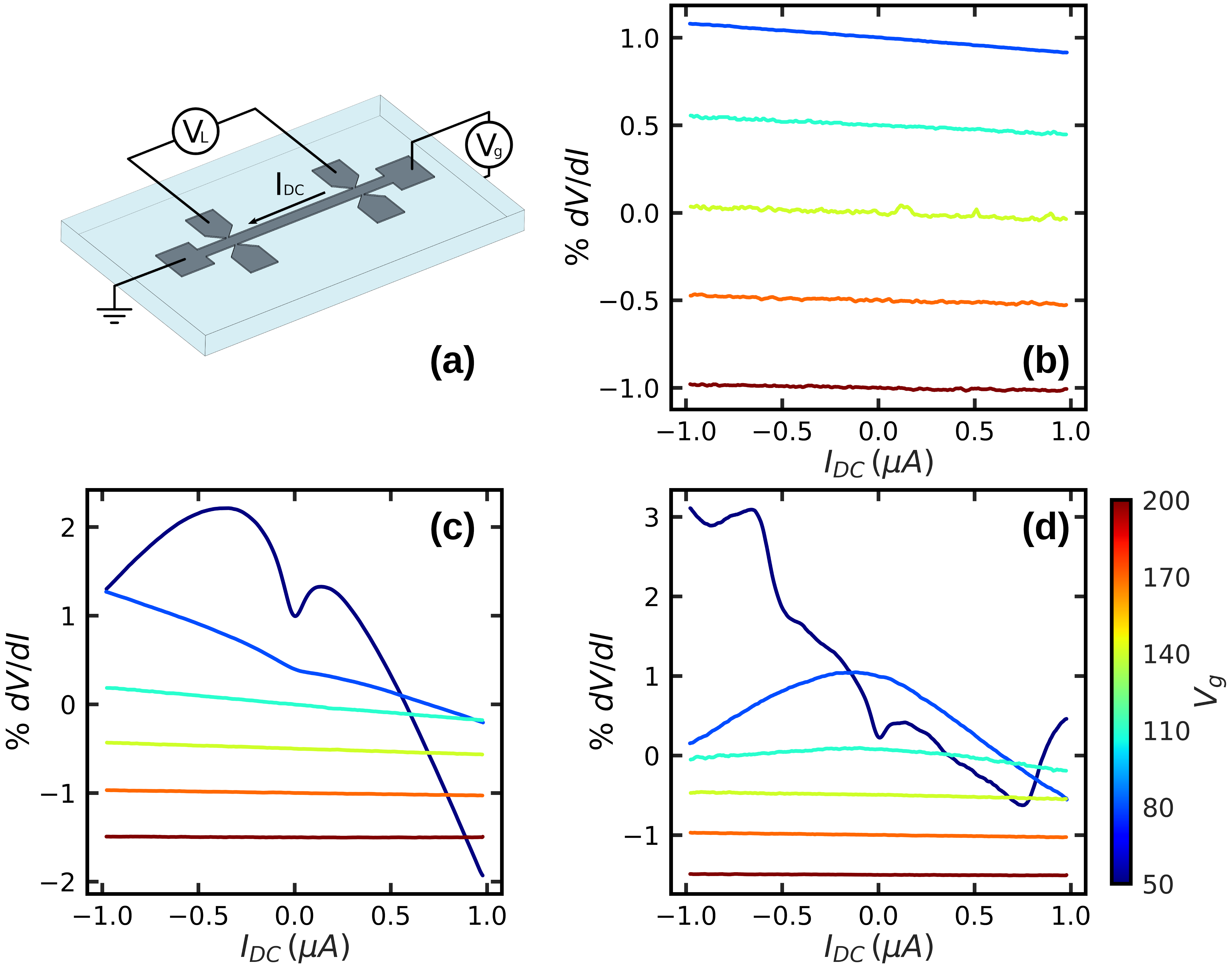}
    \caption{\label{fig:NormaldVdI}(a) Schematic diagram of the Hall bar, showing measurement probes attached for a four terminal differential resistance measurement. (b) Percent differential resistance ($dV / dI$) for the (001) sample showing a small, asymmetric deviation from constant resistance. (c) Percent ($dV / dI$) for the (110) sample showing significantly more pronounced behavior at lower gate voltages. (d) Percent ($dV / dI$) for the (111) showing the emergence of anti-symmetric peaks around a central dip. All curves have been offset by 0.5 \% for clarity.}
\end{figure}

\paragraph{}
The first measurement configuration that we employed is the standard configuration typically used to measure the longitudinal resistance of a Hall bar. Current is applied along the length of the device, and the resulting voltage measured between two probes on the same side, as shown in Fig. \ref{fig:NormaldVdI} (a). If the current-voltage (IV) characteristic of the device is linear, the differential resistance $dV/dI$ should be a constant independent of the dc bias current $I_{dc}$. While we expect a constant response, such a simple dependence is rarely observed even for canonical metals such as copper, because at some level of sensitivity and range of applied current $I_{dc}$, the $dV/dI$ does depend on $I_{dc}$. For example, if the device has a strong temperature dependence to its resistance, Joule heating due to the dc current may result in a change of $dV/dI$ on $I_{dc}$ that is symmetric in $I_{dc}$. Alternatively, if the device has a strong thermoelectric response, the thermal voltage will depend only on the magnitude of $I_{dc}$ and not its sign, resulting in a contribution to $dV/dI$ that is antisymmetric in $I_{dc}$ \cite{eom_novel_1996}. In either case, for most materials this contribution is small, especially for low dc bias currents. The presented data does not follow these expectations.

\paragraph{}
The measured differential resistance $dV/dI$ as a function of $I_{dc}$ are shown for devices fabricated on the three different crystal orientations for $V_g$ ranging from 50-200 V in Figs. \ref{fig:NormaldVdI} (b-d). Changing $V_g$ changes the charge density in the 2DEG, with higher charge densities for larger values of $V_g$. As the zero bias resistance changes by a large amount on changing $V_g$, we have plotted the data in terms of the percentage change $\Delta dV/dI$ from $I_{dc}=0$ by normalizing all curves by their zero-bias values. For large $V_g\sim 200$ V, $\Delta dV/dI$ shows essentially no dependence on $I_{dc}$ for all three crystal orientations.  As $V_g$ is decreased, $\Delta dV/dI$ starts showing an increasing variation with $I_{dc}$.  The effect is the smallest for the (001) crystal orientation ($\%dV/dI \sim 0.2$), where $\Delta dV/dI$ appears to have a mostly antisymmetric dependence on $I_{dc}$.  It is much larger for the (110) and (111) crystal orientations ($\%dV/dI \sim 3-4$), and the variation of $dV/dI$ for these orientation is also highly asymmetric in $I_{dc}$.     

\begin{figure}[H]
    \centering
    \includegraphics[width=0.75\columnwidth]{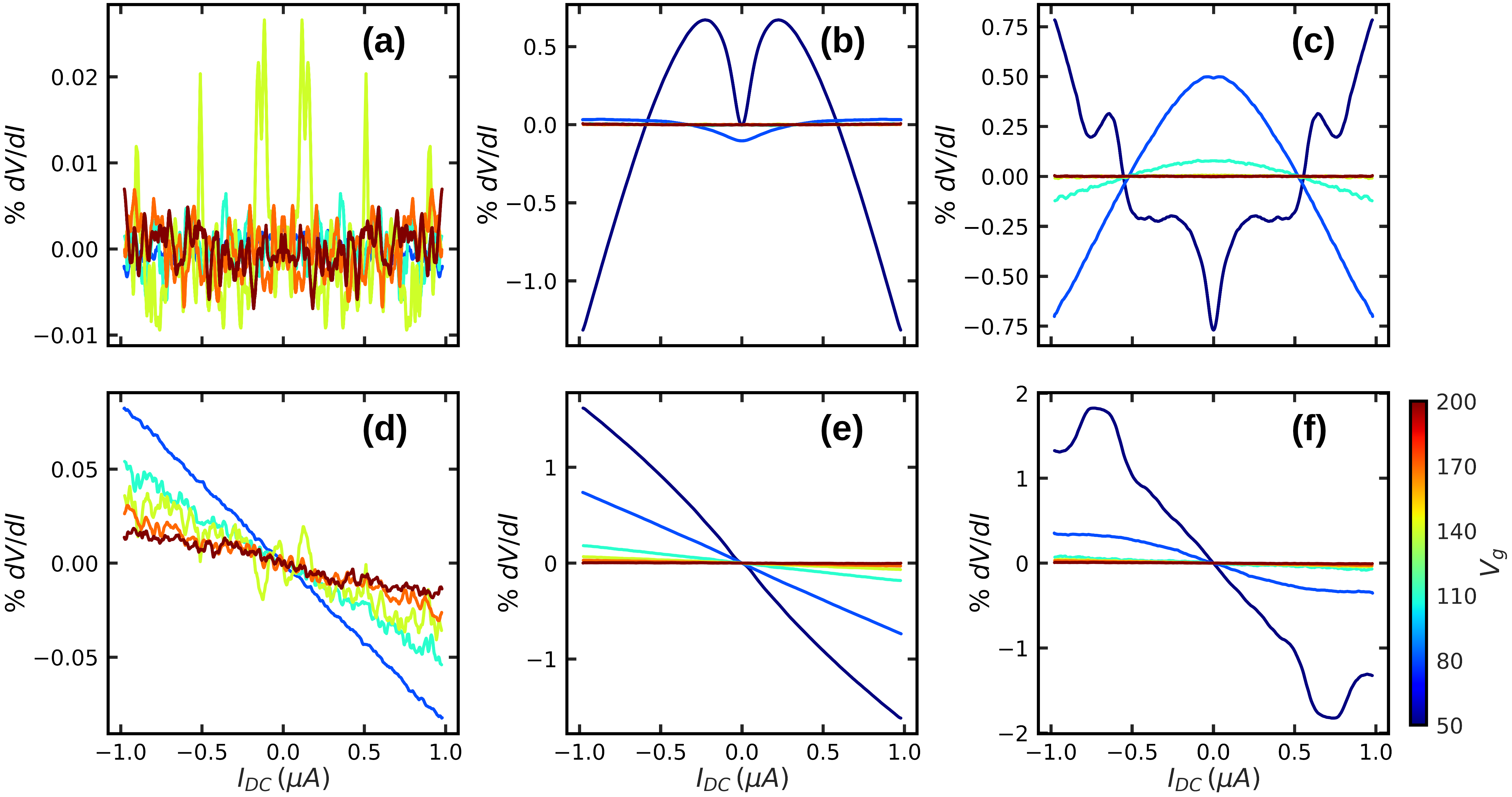}
    \caption{\label{fig:DecomposeddVdI}(a) - (c) The symmetric component of the differential resistance ($dV/dI$) of Fig. 1 as a function of gate voltage for the (001), (110), and (111) terminated samples respectively. (d) - (f) The anti-symmetric contribution of ($dV/dI$) for the same crystal orientations.}
\end{figure}

\paragraph{}
To help illuminate the symmetry contributions of $dV/dI$ with $I_{dc}$, we decompose the data of Figs. \ref{fig:NormaldVdI}(b-d) into its symmetric and anti-symmetric components. This decomposition is shown in Fig. \ref{fig:DecomposeddVdI} for all three crystal orientations. For the (001) orientation, the symmetric component is small, barely above our noise level (Fig. \ref{fig:NormaldVdI}(a)), while the antisymmetric component (Fig. \ref{fig:NormaldVdI}(d)) is small but clearly present, and grows with decreasing $V_g$.  For both the (110) and (111) devices, the components are much larger in magnitude.  We note in particular that both the symmetric and anti-symmetric components for the (111) orientation for $V_g=50$ V have a non-monotonic dependence on $I_{dc}$. 

\begin{figure}[ht]
    \centering
    \includegraphics[width=0.6\columnwidth]{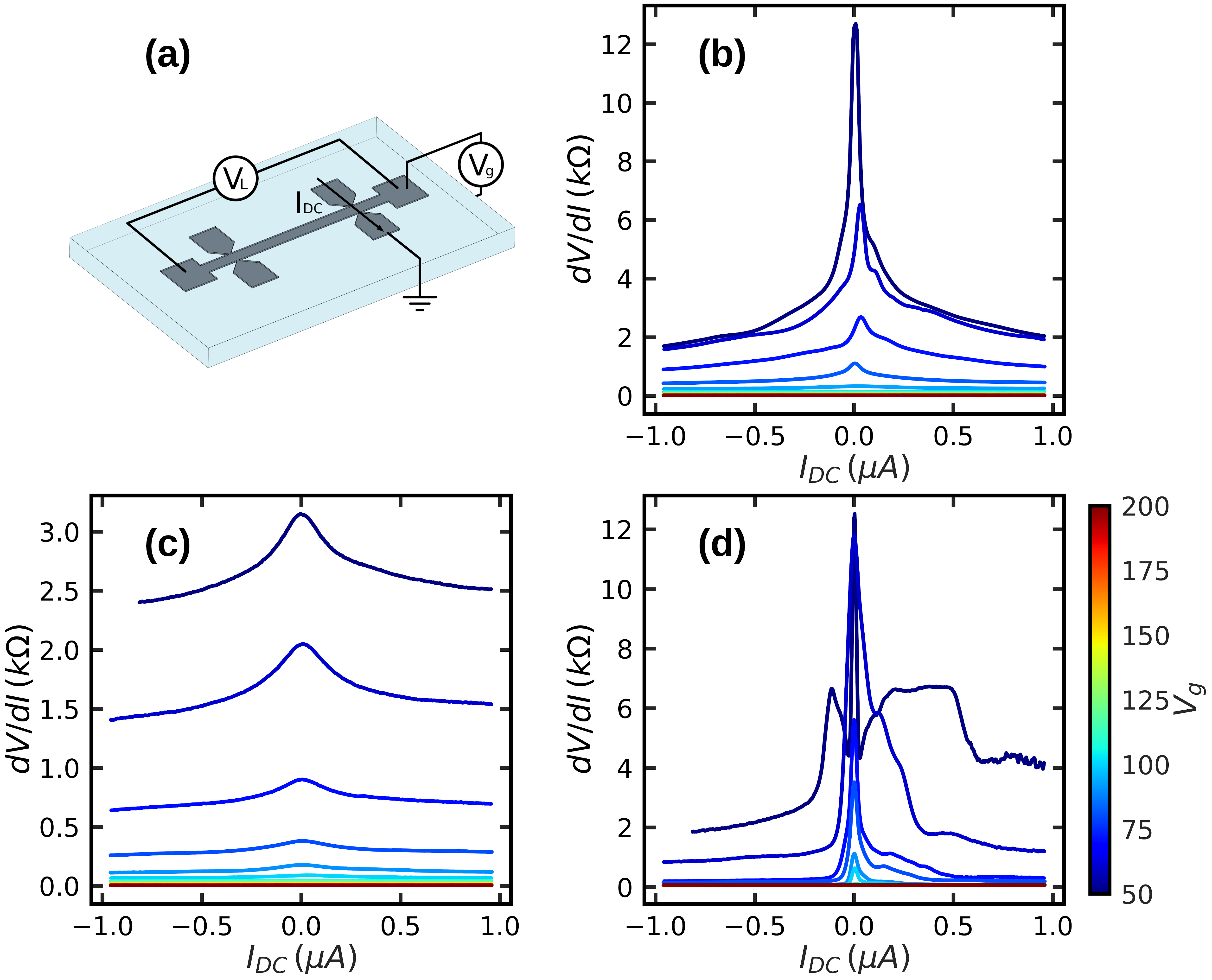}
    \caption{\label{fig:TransdVdI}(a) Schematic diagram of transverse measurement for current passed along a pair of Hall probes. (b) Transverse differential resistance ($dV / dI$) for the (001) sample showing a very large, peaked, asymmetric dependence for low values of $V_g$. (c) Transverse ($dV / dI$) for the (110) sample showing a smaller and more rounded effect. (d) Transverse ($dV / dI$) for the (111) showing an asymmetric effect complementary to the nonlocal data.}
\end{figure}

\paragraph{}
In an effort to understand the unexpected results of the normal differential resistance in the standard configuration, two other measurement configurations were studied. The first of these is shown in Fig. \ref{fig:TransdVdI} (a). In this configuration, current is passed along a set of Hall probes, and voltage is measured along the length of the Hall bar in something of an inverted Hall configuration. The resulting $dV/dI$ as a function of $I_{dc}$ for different values of $V_g$ are shown in Fig. \ref{fig:TransdVdI} (b-d) for the three crystal orientations. In the absence of an externally applied magnetic field the resulting measurement should show nothing, and for large values of $V_g$ this is indeed the case. However, the data show a large, asymmetric signal in differential resistance that increases sharply with decreasing $V_g$. The dependences are qualitatively different for the three crystal terminations. Fig. \ref{fig:TransdVdI} (b) shows the results for the (001), which has a sharply peaked feature at low values of dc current bias, and asymmetric features in current that change as a function of $V_g$. Data for the (110) sample, shown in Fig. \ref{fig:TransdVdI} (c) show a broader dependence, still asymmetric $I_{dc}$ but lacking the sharp features of the (001) device. Most striking are the data for the (111) sample, which is shown in Fig. \ref{fig:TransdVdI} (d). There, a sharp low current bias peak is complemented by more asymmetric features with stronger gate voltage dependence than even the (001) sample. In all three samples the effects drop off for higher values of dc current, but the magnitude of the low current signals extends from $\sim$ 1 - 100 k$\Omega$. 

\begin{figure}[ht]
    \centering
    \includegraphics[width=0.6\columnwidth]{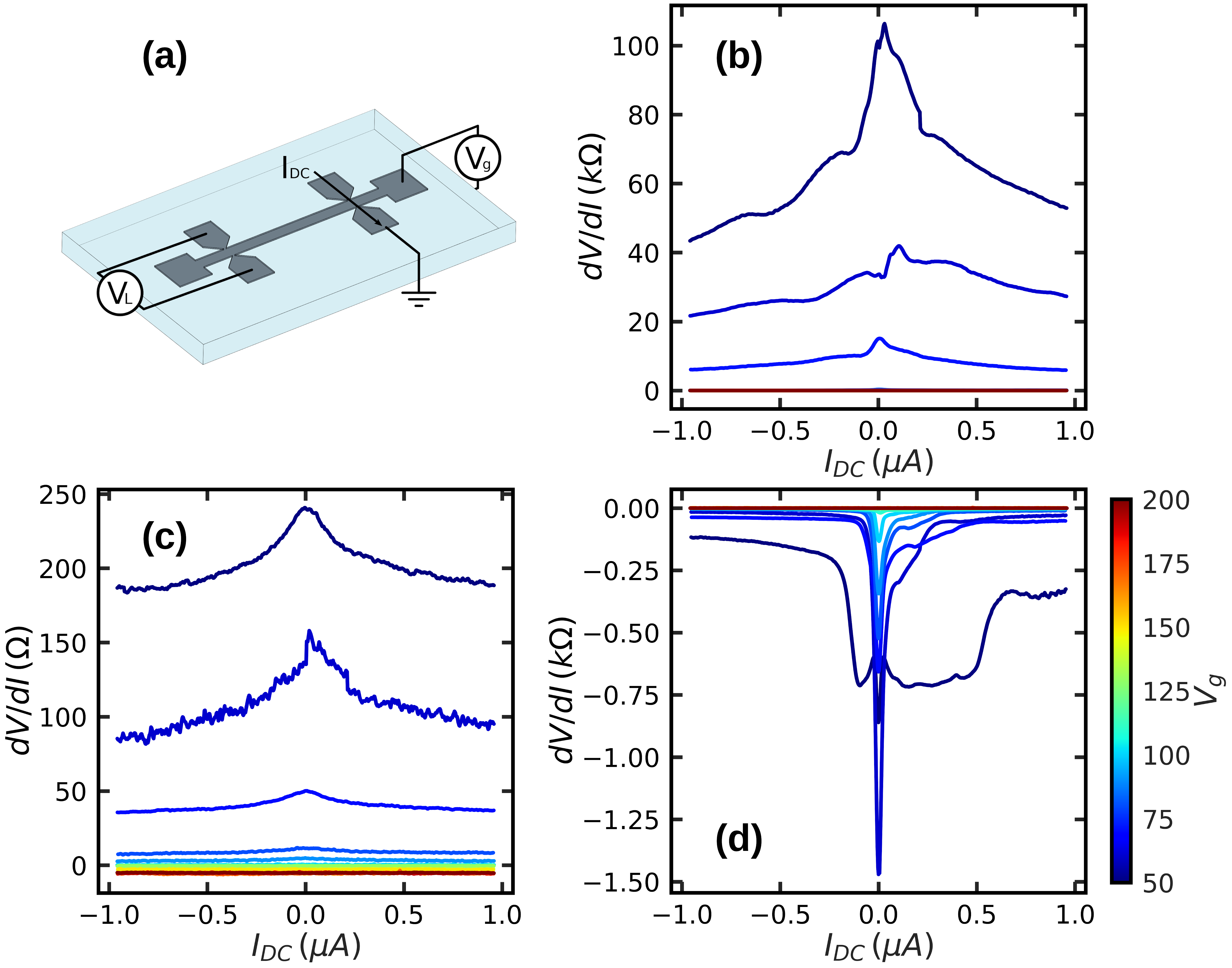}
    \caption{\label{fig:NLdVdI}(a) Schematic diagram of the Hall bar, showing measurement probes attached for a nonlocal measurement between Hall probes separated by 600 $\mu$m. (b) Nonlocal differential resistance ($dV / dI$) for the (001) sample showing a very large, asymmetric dependence for low values of $V_g$. (c) Nonlocal ($dV / dI$) for the (110) sample showing a smaller but still asymmetric effect. (d) Nonlocal ($dV / dI$) for the (111) showing a negative and strongly asymmetric effect that persists to much higher values of $V_g$ than in the (001) or (110).}
\end{figure}

\paragraph{}
Results from the third measurement configuration are perhaps the most surprising.  This configuration is shown in Fig. \ref{fig:NLdVdI} (a).This is a nominally nonlocal configuration in which current is passed along one set of voltage probes across the Hall bar, and voltage is measured between the other pair of voltage probes. To remind the reader, in our devices, these second set of Hall probes are at a distance of 600 $\mu$m from the first set of Hall probes, and so at the same distance from the current path. Such a measurement configuration is used to detect spin Hall effects in systems that have intrinsic spin charge conversion \cite{jin_nonlocal_2017}, but with a much shorter distance between the injection and detection probes. Those measurements are also conducted in an applied magnetic field, parallel to the plane of the Hall bar. No such field is applied here. As was seen in the second measurement configuration, the measurements for this configuration give a large, asymmetric signal that increases sharply with decreasing $V_g$. These large, asymmetric signals are most prominent in the (001) sample, as seen in Fig. \ref{fig:NLdVdI} (b), reaching $\sim$ 100 k$\Omega$, or the equivalent of nearly half a volt of potential at the largest measured dc currents. The corresponding data for the (110) sample is shown in Fig. \ref{fig:NLdVdI} (c), reaching only a few hundred ohms at the lowest gate voltages, but still showing a central resistance peak and a modest asymmetry in current. Once again the signal for the (111) sample, shown in Fig. \ref{fig:NLdVdI} (d), shows the most prominent features of all the samples with a very sharp central peak and multiple asymmetric features strongly dependent on $V_g$. Again, we emphasize that the probes across which the voltage is measured are separated from the nominal current path by 600 $\mu$m, so observation of any response would be surprising, let alone one of such magnitude.

\begin{figure}[ht]
    \centering
    \includegraphics[width=0.6\columnwidth]{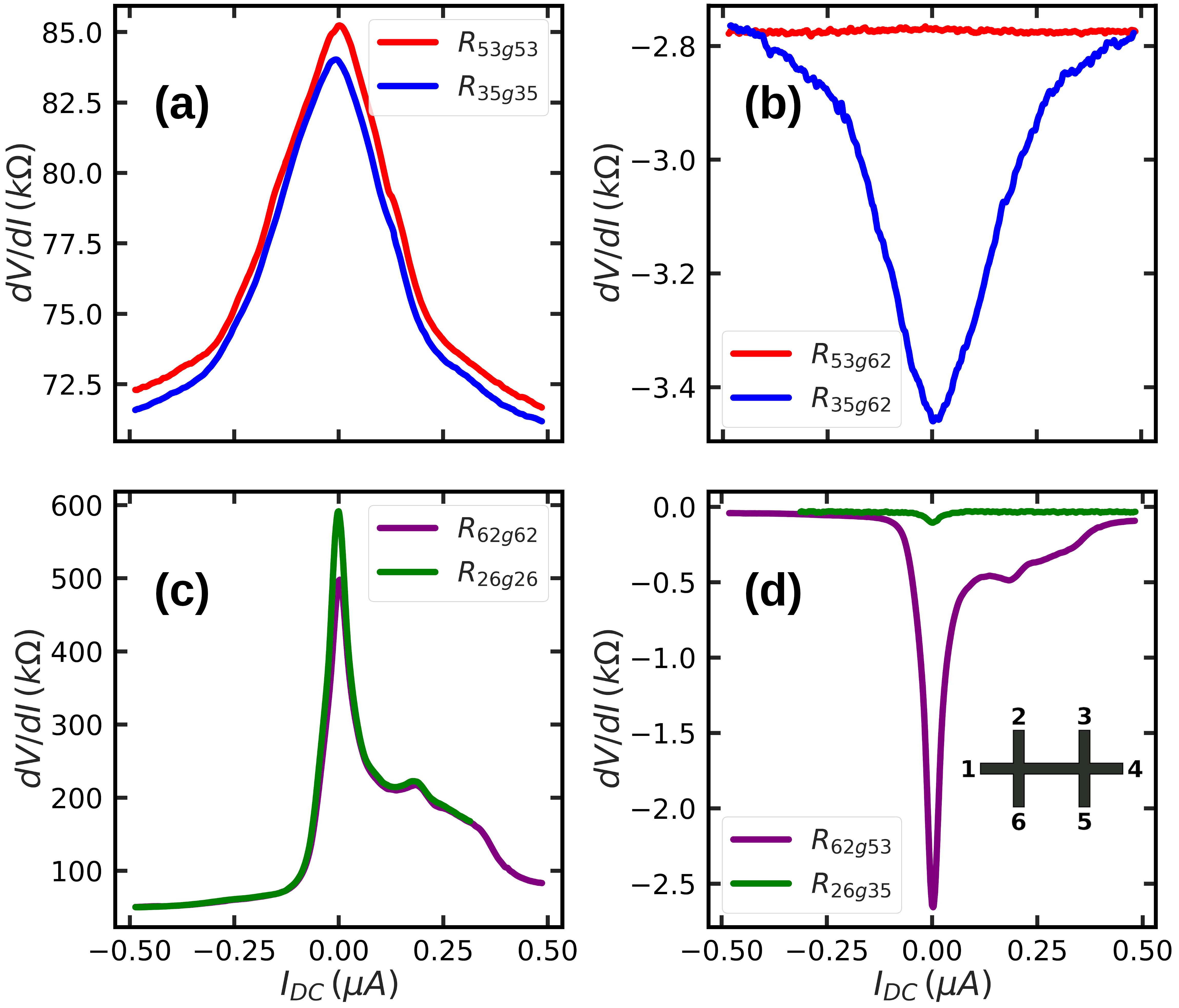}
    \caption{\label{fig:ProbeChange}(a) Two terminal measurements demonstrating differential resistance for current passed from probes 3 to 5, and from 5 to 3. (b) Differential resistance measured nonlocally between probes 6 and 2 for the two current configurations described in (a). (c) Psuedo - two terminal measurements for current passed from probes 6 to 2, and from 2 to 6. (d) Nonlocal differential resistance for current passed from 6 to 2 and the polarity complement. Inset: Hall bar schematic with probes labeled to describe connections for the reported data. }
\end{figure}

\paragraph{}
In an effort to determine the origin of this large nonlocal signal, we performed a series of probe switching experiments in order to compare the results from different 4-probe measurement configurations. In order to keep track of the different configurations, we will use the notation $R_{ij,kl}$, where $i,j$ label the probes used to source and sink the measurement current ($I+,I-$), and $k,l$ label the leads used to measure the resulting voltage ($V+,V-$). As a consistency check, we also performed two-terminal measurements, which are denoted by $R_{ij,ij}$. Finally, the $I-$ lead in any measurement configuration is grounded, and to emphasize this, we add the symbol $g$ to denote which lead is grounded. Fig. \ref{fig:ProbeChange} shows the results of this examination just for the (111) sample at $V_g$ = 50 V. The numbering scheme used is shown in the inset schematic for Fig. \ref{fig:ProbeChange} (d). 

\paragraph{}
To start, a set of consistency checks were performed by measuring the two-terminal differential resistance in two different configurations. Fig. \ref{fig:ProbeChange} (a) shows the two terminal differential resistance for current passed from Hall probe 3 to 5 and compares it to the same measurement but with current passed from 5 to 3, but reflected in dc current for ease of comparison. While this approach adds in extra contributions to the measured resistance from the cryostat wiring, it demonstrates the consistency of the measurement and makes a foundation of comparison for further measurements. This is important, because Fig. \ref{fig:ProbeChange} (b) shows a pair of curves taken simultaneously to the curves shown in Fig. \ref{fig:ProbeChange} (a), but with the voltage measured along the other set of Hall probes, in a nonlocal configuration. Both nonlocal measurements in Fig. \ref{fig:ProbeChange} (b) are taken with the same voltage connections but show dramatically different results for a reversal of the current direction. For one direction, there is a large signal that reflects the differential resistance seen in Fig. \ref{fig:ProbeChange} (a), albeit with much smaller amplitude, while in the other direction, the signal vanishes.  Reversing the voltage leads while keeping the current leads the same reverses the sign of the signal, as expected.

\paragraph{}
A similar set of checks was completed on the other set of Hall probes, as can be seen in Fig. \ref{fig:ProbeChange} (c). Here the probes used for the two-terminal measurements are 6 and 2. The shape of the two curves are nearly identical, and again form a basis of comparison for further nonlocal tests. Fig. \ref{fig:ProbeChange} (d) shows a comparison between probes where both the polarity of the current and the polarity of the voltage probes have been swapped, demonstrating again the change in signal dependent on changes in current injection, and the fact that the nonlocal signal inherits the shape of its current dependence from the injection lead. 

\paragraph{}
That the presence or absence of a signal depends on the probes used for sourcing and sinking the dc current is puzzling, because when the signal is present, as for $R_{35g,62}$ in Fig. \ref{fig:ProbeChange} (b), one sees it for both positive and negative $I_{dc}$. Another possibility is perhaps that the presence or absence of the signal depends on which probe is grounded. For example, referring to the data of Fig. \ref{fig:ProbeChange} (b), it might be that the signal disappears if lead 3 is grounded, but not lead 5, which would potentially point to an artifact of our measurement.  However, the two-terminal measurements shown in Figs. \ref{fig:ProbeChange} (a) and (c) show a large signal regardless of which lead is grounded, so we can discount the possibility that the manner in which we ground the sample is what is giving rise to the anomalous behavior. 

\paragraph{}
What is the possible origin of the effects we observe?  We note first that the anomalous behavior we observe in all three configurations appears to be related, so that any explanation must account for the full set of observations. We noted earlier two possibilities that may give rise to an asymmetric differential resistance in our first measurement configuration.  These were the temperature dependent resistance, which would give rise to a contribution to $dV/dI$ that is symmetric in $I_{dc}$, and a thermoelectric response that would give rise to a contribution to $dV/dI$ that is antisymmetric in $I_{dc}$. Our measurements in the last (nonlocal) measurement configuration discount both these possibilities. Regardless of whether the device has a strong temperature dependent resistance or not, one does not expect a nonlocal contribution on voltage probes placed 600 $\mu$m away. Similarly, a thermoelectric contribution should not depend on the sign of the current, and should certainly not disappear if the current source and sink are swapped. A strong dependence on the direction of current injection combined with the observed asymmetry brings to mind studies of mesoscopic effects in nanowires \cite{eom_asymmetric_1996}, but these samples are comparatively massive, and the signals significantly larger than in mesoscopic systems. Another potential explanation is that at lower gate voltages, where the sample becomes more resistive, the sample is becomes highly disordered and the transport percolative, so that current paths through the sample may diverge significantly.  Finally, given that KTO is known to have strong spin-orbit interactions and spin-momentum locking, it may be that the transport is highly chiral. However, it would still be hard to fathom that these latter two explanations can give rise to measurable effects over length scales of hundreds of microns.    

\paragraph{}
While the root cause of the anomalous differential resistances shown here cannot be determined from the data, the striking magnitude and asymmetric features remain important signatures to recognize. Data taken through a variety of probe combinations on a standard Hall bar geometry show several remarkable trends, most prominently at the lowest measured gate voltages. These behaviors are large in magnitude, rising to hundreds of thousands of ohms for some combination of crystal terminations and measurement geometries. They have a large asymmetric dependence on the current direction with reproducible features that track between different measurement geometries. All of these points demand further testing to seek the root cause of these behaviors, potentially using smaller devices with more strictly defined conducting paths. 


\medskip
\textbf{Supporting Information} \par 
Supporting Information is available from the Wiley Online Library or from the author.

\medskip
\textbf{Acknowledgements} \par 
This work was supported by the U.S. Department of Energy, Basic Energy Sciences, under Award No. DE-FG02-06ER46346. This work also made use of the NUFAB facility of Northwestern University’s NUANCE Center, which has received support from the SHyNE Resource (NSF ECCS-2025633), the IIN, and Northwestern’s MRSEC program (NSF DMR-1720139). Additional equipment support was provided by DURIP grant W911NF-20-1-0066.

\medskip

%
\bibliographystyle{MSP}
\bibliography{PSS_RRL.bib}








\end{document}